\documentclass[aps,prc,twocolumn,showpacs,superscriptaddress,longbibliography,nofootinbib,floatfix,10pt]{revtex4-1}

\usepackage{bm}
\usepackage{dcolumn}% Align table columns on decimal
\usepackage{amssymb}
\usepackage{amsmath}
\usepackage{graphicx}
\usepackage{dcolumn}% Align table columns on decimal point
\usepackage{bm}
\usepackage{xcolor}
\usepackage{fix-cm}
\usepackage{mathptmx} % rm & math
\usepackage[T1]{fontenc}
\usepackage[colorlinks,citecolor=blue,linkcolor=blue,anchorcolor=blue,filecolor=blue,urlcolor=blue]{hyperref}
\usepackage{multirow}
\makeatletter
\def\NAT@def@citea{\def\@citea{\NAT@separator}}
\makeatother

\begin{document}

%\title{The role of tensor force on heavy-ion interaction potential deduced from density-constraint time-dependent Hartree-Fock approach}
\title{The influence of tensor force on the microscopic heavy-ion interaction potential}

\author{Lu Guo} \email{luguo@ucas.ac.cn}
\affiliation{School of Nuclear Science and Technology, University of Chinese Academy of Sciences, Beijing 100049, China}
\affiliation{Institute of Theoretical Physics, Chinese Academy of Sciences, Beijing 100190, China}
\author{K. Godbey} \email{kyle.s.godbey@vanderbilt.edu}
\affiliation{Department of Physics and Astronomy, Vanderbilt University, Nashville, TN 37235, USA}
\author{A. S. Umar} \email{umar@compsci.cas.vanderbilt.edu}
\affiliation{Department of Physics and Astronomy, Vanderbilt University, Nashville, TN 37235, USA}

\date{\today}

\begin{abstract}
\edef\oldrightskip{\the\rightskip}
\begin{description}
        \rightskip\oldrightskip\relax
        \setlength{\parskip}{0pt}
\item[Background]
The tensor interaction is known to play an important role in the nuclear structure studies of exotic nuclei.
However, most microscopic studies of low-energy nuclear reactions
neglect the tensor force, resulting in a lack of knowledge concerning the effect of the tensor force on heavy-ion collisions.
An accurate description of the heavy-ion interaction
potential is crucial for understanding the microscopic mechanisms of heavy-ion fusion dynamics.
Furthermore, the building blocks of the heavy-ion interaction potential in terms of the ingredients of the
effective nucleon-nucleon interaction provides the physical underpinnings for connecting the theoretical
results with experiment.
The tensor force has never been
incorporated for calculating the nucleus-nucleus interaction potential.
\item[Purpose]
The theoretical study of the influence of the tensor force on heavy-ion interaction potentials is required to further our
understanding of the microscopic mechanisms entailed in fusion dynamics.
\item[Method]
The full Skyrme tensor force is implemented into the static Hartree-Fock and dynamic density-constrained time-dependent Hartree-Fock
(DC-TDHF) theory
to calculate both static (frozen density) and dynamic microscopic interaction potentials for reactions involving exotic and stable nuclei.
\item[Results]
The static potentials are found to be systematically higher than the dynamical results, which are attributed to the microscopic dynamical effects included in TDHF.
We also show that the dynamical potential barriers vary more significantly by the inclusion of tensor force than the static barriers.
The influence of isoscalar and isovector tensor terms is also investigated with the T$IJ$ set of forces.
For light systems, the tensor force is found to have an imperceptible effect on the nucleus-nucleus potential. However, for medium and heavy spin-unsaturated reactions,
the potentials may change from a fraction of an MeV to almost 2~MeV by the inclusion of tensor force, indicating a strong impact of the tensor force on sub-barrier fusion.
\item[Conclusions]
The tensor force could indeed play a large role in the fusion of nuclei, with spin-unsaturated systems seeing a systematic increase in ion-ion barrier height and width. This fusion hindrance is partly due to static, ground state effects from the inclusion of the tensor force, though additional hindrance appears when studying nuclear dynamics.
\end{description}
\end{abstract}

%\pacs{}
%\keywords{}

\maketitle

\section{Introduction}
\label{introduction}

With the increasing availability of radioactive ion beams, the study of structure and reactions of exotic nuclei is
one of the most fascinating research areas in nuclear physics~\cite{Balantekin2014_MPLA29-1430010}.
The exotic nuclei display distinct features from those seen in typical stable nuclei, which is attributed partly to the unique characteristics of the nucleon-nucleon interaction.
The tensor interaction between nucleons is one such characteristic and is well known to be important in nuclear structure
properties~\cite{Lesinski2007_PRC76-014312}, e.g.,
the shell evolution of exotic nuclei~\cite{Otsuka2006_PRL97-162501},
spin-orbit splitting~\cite{Colo2007_PLB646-227}, and Gamow-Teller and charge exchange spin-dipole excitations~\cite{Bai2010_PRL105-072501}.
However, its role in low-energy nuclear reactions is poorly understood as the tensor force has been neglected in most reaction dynamics calculations.
In particular, regarding nuclear dynamics, the tensor force changes not only the spin-orbit splitting but also the intrinsic excitations which may give rise to dynamical effects which are more complicated than those arising from simple shell evolution.
The study of the effects of the tensor force on heavy-ion fusion dynamics will lead to a better understanding of the effective nucleon-nucleon interaction and of the correlations present in these many-body systems.

The study of heavy-ion interaction potentials is of fundamental importance for above barrier and sub-barrier fusion reactions~\cite{Back2014_RMP86-317}.
In general, two categories of theoretical approaches are used for
calculating ion-ion potentials. In the first category, phenomenological models such as the Bass model~\cite{Bass1974_NPA231-45}, the proximity potential~\cite{Randrup1978_PLB77-170,Seiwert1984_PRC29-477}, the double-folding potential~\cite{Satchler1979_PR55-183,Rhoades-Brown1983_PRL50-1435}, and driven potential from dinuclear system model~\cite{Adamian2004_PRC69-044601,
Wang2012_PRC85-041601,Zhu2016_PRC93-064610,Bao2016_PRC93-044615,Feng2018_PRC95-024615} could be mentioned.
 Although these methods have been successful in explaining particular aspects of reaction
data~\cite{Randrup1978_NPA307-319,Fazio2004_EPJA19-89}, the uncertainty of macroscopic parameters and the lack of microscopic origins restrict their predictive power and may obscure the underlying physical processes.
Second category contains the semi- and fully microscopic approaches to obtain potentials by including the interactions of the constituents~\cite{Moller2004_PRL92-072501,Guo2004_NPA740-59,Guo2005_PRC71-024315,Guo2007_PRC76-034317,Lu2014_PRC89-014323}.
One common  assumption used in many of the semi-microscopic calculations is that of the frozen density or sudden approximation~\cite{Brueckner1968_PR173-944}, in which the nuclear densities are unchanged during the computation of the nucleus-nucleus potential as a function of internuclear distance.
This approximation may result in an unphysical potential at deep sub-barrier energies, where the inner turning point of the interaction potential corresponds to large nuclear overlap.
Various remedies have been developed to address this issue within the confines of the coupled-channels approach~\cite{Misicu2006_PRL96-112701,Ichikawa2007_PRC75-057603}.
In other microscopic approaches, such as the constrained mean-field methods, although the nuclear densities are allowed for the rearrangement, the potential energy path is obtained by the static adiabatic approximation, thus ignoring the dynamical effects.

In recent years we have developed the density-constrained time-dependent Hartree-Fock (DC-TDHF) approach for calculating heavy-ion interaction
potentials, which naturally incorporate all of the dynamical effects included in the time-dependent Hartree-Fock (TDHF) description of the collision process~\cite{Umar2006_PRC74-021601}.
These effects include nucleon transfer, couplings between the collective motion and intrinsic degrees of freedom, neck formation, internal excitations, and deformation effects to all orders.
The method is based on the TDHF evolution of the nuclear dynamics coupled with density-constrained (DC) Hartree-Fock (HF) calculations to obtain the ion-ion potential.
In contrast to other mean-field based microscopic methods, the DC-TDHF approach doesn't need
to introduce external constraining operators which assume that the collective motion is confined to the constrained phase space. That means that the many-body system
selects its evolutionary path by itself following the microscopic dynamics. We have applied this method for a wide range of reactions ~\cite{Umar2006_PRC74-024606,Umar2006_PRC74-061601,Umar2008_PRC77-064605,Umar2008_EPJA37-245, Umar2009_PRC80-041601,Oberacker2010_PRC82-034603,Keser2012_PRC85-044606,Umar2012_PRC85-055801,
Umar2014_PRC89-034611,Godbey2017_PRC95-011601} and found reasonable agreement between the measured fusion cross sections and the DC-TDHF results.
To our knowledge, neither the phenomenological nor microscopic methods for calculating ion-ion potential include the tensor
force between nucleons, which is an important component of the nuclear force. Our work is the first attempt to investigate the effect induced by the tensor
force on heavy-ion interaction potentials.

The TDHF approach is a well-defined framework and provides a useful foundation for a fully microscopic many-body theory.
Quantum effects are considered, which is essential for the manifestation of shell effects during the collision dynamics.
Recently, the effect of the tensor force in heavy-ion collisions has been studied using direct TDHF calculations~\cite{Fracasso2012_PRC86-044303,Dai2014_SciChinaPMA57-1618,Stevenson2016_PRC93-054617,Shi2017_NPR34-41,Guo2018_PLB782-401}.
Furthermore, the TDHF approach provides a deeper understanding of nuclear dynamics, as seen in recent applications to fusion~\cite{Simenel2004_PRL93-102701,Umar2009_PRC80-041601,Oberacker2010_PRC82-034603,Guo2012_EPJWoC38-09003,
Keser2012_PRC85-044606,Umar2012_PRC85-055801,Simenel2013_PRC88-024617,
Umar2014_PRC89-034611,Washiyama2015_PRC91-064607,Tohyama2016_PRC93-034607,Godbey2017_PRC95-011601,Simenel2017_PRC95-031601},
quasifission~\cite{Golabek2009_PRL103-042701,Oberacker2014_PRC90-054605,Umar2015_PRC92-024621,Umar2016_PRC94-024605,Yu2017_SciChinaPMA60-092011}, transfer reactions~\cite{Washiyama2009_PRC80-031602,Simenel2010_PRL105-192701,Simenel2011_PRL106-112502,Scamps2013_PRC87-014605,Sekizawa2013_PRC88-014614,
Wang2016_PLB760-236,Sekizawa2016_PRC93-054616,Sekizawa2017_PRC96-041601},
fission~\cite{Simenel2014_PRC89-031601,Scamps2015_PRC92-011602,Goddard2015_PRC92-054610,Goddard2016_PRC93-014620,Bulgac2016_PRL116-122504,Tanimura2017_PRL118-152501},
and deep inelastic collisions~\cite{Maruhn2006_PRC74-027601,Guo2007_PRC76-014601,Guo2008_PRC77-041301,Iwata2011_PRC84-014616,Dai2014_PRC90-044609,Dai2014_SciChinaPMA57-1618,
Stevenson2016_PRC93-054617,Guo2017_EPJWoC163-00021,Shi2017_NPR34-41,Umar2017_PRC96-024625}.
%and resonance dynamics~\cite{Simenel2001_PRL86-2971,
%Simenel2003_PRC68-024302,Maruhn2005_PRC71-064328,Nakatsukasa2005_PRC71-024301,Umar2005_PRC71-034314,Reinhard2007_EPJA32-19,Simenel2009_PRC80-064309,
%Fracasso2012_PRC86-044303,Avez2013_EPJA49-76,Ebata2014_PRC90-024303}.
For recent reviews see Refs.~\cite{Simenel2012_EPJA48-152,Nakatsukasa2016_RMP88-045004,Simenel2018_PPNP}.

This article is organized as follows. In Sec.~\ref{theory}, we summarize the theoretical formalism of Skyrme energy functional with the tensor force included and the TDHF and DC-TDHF approaches.
Section~\ref{discuss} presents the systematic analysis of the impact of the tensor force on heavy-ion interaction potentials.
A summary is given in Sec.~\ref{summary}.

\section{Theoretical framework}
\label{theory}

Despite the wide application of the TDHF approach, various assumptions and approximations that may affect the TDHF results have been employed in the past.
This led to an occasional imperfect or even incorrect reproduction of experimental data.
To remedy these problems a considerable theoretical and computational effort has been undertaken for increased numerical accuracy
and improved effective interactions.
For instance, the inclusion of the spin-orbit interaction solved an early conflict between TDHF predictions and experimental observations~\cite{Umar1986_PRL56-2793,Umar1989_PRC40-706} and turned out to play an important role in fusion and dissipation dynamics~\cite{Maruhn2006_PRC74-027601,Dai2014_PRC90-044609}. In recent years it has become feasible
to perform TDHF calculations on a three-dimensional Cartesian grid without any symmetry restrictions and with accurate numerical methods. In
addition, the quality of energy density functional (EDF) has been substantially improved. The time-odd
terms in particular have shown to be non-negligible in heavy-ion collisions~\cite{Umar2006_PRC73-054607}. However, there are still important components of the basic theory that have not yet been fully implemented, such as the tensor force between nucleons. In order to study the role of tensor force
in heavy-ion interaction potential, we incorporate the full tensor force into the microscopic TDHF
and DC-TDHF approaches.

\subsection{Full Skyrme energy functional}

Most TDHF calculations employ the Skyrme effective interaction~\cite{Skyrme1956_PM1-1043}, in which the two-body tensor force was proposed in its original form as
\begin{align}
\begin{split}
v_T&=\dfrac{t_\mathrm{e}}{2}\bigg\{\big[3({\sigma}_\mathrm{1}\cdot\mathbf{k}')({\sigma}_\mathrm{2}\cdot\mathbf{k}')-({\sigma}_\mathrm{1}\cdot{\sigma}_\mathrm{2})\mathbf{k}'^{\mathrm{2}}\big]\delta(\mathbf{r}_\mathrm{1}-\mathbf{r}_\mathrm{2})\\
&+\delta(\mathbf{r}_\mathrm{1}-\mathbf{r}_\mathrm{2})\big[3({\sigma}_\mathrm{1}\cdot\mathbf{k})({\sigma}_\mathrm{2}\cdot\mathbf{k})-({\sigma}_\mathrm{1}\cdot{\sigma}_\mathrm{2})\mathbf{k}^\mathrm{2}\big]\bigg\}\\
&+t_\mathrm{o}\bigg\{3({\sigma}_\mathrm{1}\cdot\mathbf{k}')\delta(\mathbf{r}_\mathrm{1}-\mathbf{r}_\mathrm{2})({\sigma}_\mathrm{2}\cdot\mathbf{k})-({\sigma}_\mathrm{1}\cdot{\sigma}_\mathrm{2})\mathbf{k}'
\delta(\mathbf{r}_\mathrm{1}-\mathbf{r}_\mathrm{2})\mathbf{k}\bigg\}.
\end{split}
\end{align}
The coupling constants $t_\textrm{e}$ and $t_\textrm{o}$ represent the strengths of triplet-even and
triplet-odd tensor interactions, respectively.  The operator $\mathbf{k}=\frac{1}{2i}(\nabla_1-\nabla_2)$ acts on the right and
$\mathbf{k}'=-\frac{1}{2i}(\nabla'_1-\nabla'_2)$ acts on the left.

It is natural to represent the total energy of the system
\begin{equation}
E=\int d^3r  {\cal H}(\rho, \tau, {\mathbf{j}}, {\textbf{s}}, {\textbf{T}}, {\textbf{F}}, J_{\mu\nu}; {\mathbf{r}})
\label{Energy}
\end{equation}
in terms of the energy functional. The functional is composed by
the number density $\rho$, kinetic density $\tau$, current density ${\mathbf{j}}$, spin density ${\mathbf{s}}$, spin-kinetic density ${\mathbf{T}}$,
the tensor-kinetic density {\textbf{F}}, and spin-current pseudotensor density $J$~\cite{Stevenson2016_PRC93-054617}. The full version of Skyrme EDF is expressed as
\begin{align}
\label{EDFH}
\begin{split}
\mathcal{H}&=\mathcal{H}_0+\sum_{\rm{t=0,1}}\Big\{A_{\rm{t}}^{\rm{s}}\mathbf{s}_{\rm{t}}^2+(A_{\rm{t}}^{\Delta{s}}+B_{\rm{t}}^{\Delta{s}})
\mathbf{s}_{\rm{t}}\cdot\Delta\mathbf{s}_{\rm{t}}+B_{\rm{t}}^{\nabla s}(\nabla\cdot \mathbf{s}_{\rm{t}})^2 \\
&+B_{\rm{t}}^{F}\big(\mathbf{s}_{\rm{t}}\cdot
\mathbf {F}_{\rm{t}}-\frac{1}{2}\big(\sum_{\mu=x}^{z}J_{\rm{t}, \mu\mu}\big)^2-\frac{1}{2}\sum_{\mu, \nu=x}^{z}J_{\rm{t}, \mu\nu}J_{\rm{t}, \nu \mu}\big)\\
&+(A_{\rm{t}}^{\rm{T}}+B_{\rm{t}}^{\rm{T}})\big(\mathbf{s}_{\rm{t}}\cdot\mathbf{T}_{\rm{t}}-
\sum_{\mu,\nu=x}^{z}J_{\rm{t},\mu\nu}J_{\rm{t},\mu\nu}\big)\Big\},
\end{split}
\end{align}
where $\mathcal{H}_0$ is the simplified functional used in the Sky3D TDHF code~\cite{Maruhn2014_CPC185-2195} and most TDHF calculations.
The terms containing the coupling constants $A$ arise from the Skyrme central force and those with $B$ from the tensor force. The definitions of both $A$ and $B$
can be found in Refs.~\cite{Lesinski2007_PRC76-014312,Davesne2009_PRC80-024314}. All the time-even and time-odd terms in Eq.~(\ref{EDFH}) have been implemented numerically in the mean-field Hamiltonians of the HF, TDHF and
DC-TDHF approaches.
As pointed out in Refs.~\cite{Lesinski2007_PRC76-014312,Stevenson2016_PRC93-054617}, the terms containing the gradient of spin density may cause spin instability in both nuclear structure and reaction studies, hence the terms of $\mathbf{s}_{\rm{t}}\cdot\Delta\mathbf{s}_{\rm{t}}$ and $(\nabla\cdot \mathbf{s}_{\rm{t}})^2$ have been turned off in all calculations.

\subsection{TDHF approach}

Given a many-body Hamiltonian, the action can be constructed as
\begin{equation}
S=\int_{t_1}^{t_2} dt \langle \Phi(\textbf{r}, t)|H-i\hbar \partial_t|\Phi(\textbf{r}, t)\rangle ,
\end{equation}
where $\Phi$ is the time-dependent many-body wave function.
In TDHF approach the many-body wave function $\Phi(\mathbf{r}, t)$ is approximated as a single time-dependent Slater determinant composed of an antisymmetrized product
of the single particle states $\phi_{\rm{\lambda}}(\mathbf{r}, t)$
\begin{equation}
\Phi(\mathbf{r}, t)=\frac{1}{\sqrt{N!}}\textrm{det}\{\phi_{\rm{\lambda}}(\mathbf{r}, t)\},
\end{equation}
and this form is kept at all times in the dynamical evolution.
By taking the variation of the action with respect to the single-particle wave functions, the set of nonlinear coupled TDHF equations in the multidimensional
space-time phase space
\begin{equation}
i\hbar \frac{\partial}{\partial_t}\phi_{\rm{\lambda}}(\mathbf{r}, t)=h\phi_{\rm{\lambda}}(\mathbf{r}, t)
\end{equation}
yields the most probable time-dependent mean-field path, where $h$ is the HF single-particle Hamiltonian.
The set of nonlinear TDHF equations has been solved on three-dimensional
coordinate space without any symmetry restrictions and with modern, accurate numerical methods~\cite{Umar2006_PRC73-054607,Maruhn2014_CPC185-2195}.

\subsection{Dynamical potential from DC-TDHF approach}

Since TDHF theory describes the collective motion of fusion dynamics in terms of semi-classical trajectories, the sub-barrier quantum
tunneling of the many-body wave function can not be included. Consequently, direct TDHF calculations can not be used to describe sub-barrier fusion.
At present, all sub-barrier fusion calculations assume that there exists an ion-ion potential which depends on the internuclear distance.
The microscopic DC-TDHF approach~\cite{Umar2006_PRC74-021601} is employed to extract the nucleus-nucleus potential from the TDHF time evolution of the dinuclear system.
In this approach, at certain time during the evolution, the instantaneous TDHF density is used to perform a static HF energy minimization
\begin{equation}
\delta \langle \Psi_{\rm DC}|H-\int d^3r \lambda(\textbf{r})\rho(\textbf{r})|\Psi_{\rm DC}\rangle=0,
\end{equation}
by constraining the proton and neutron densities to be equal to the instantaneous TDHF densities. Since we are constraining the total density,
all moments are simultaneously constrained. DC-TDHF calculations give the adiabatic reference state for a given TDHF state, which is the Slater
determinant with the lowest energy for a given density.
The minimized energy
\begin{equation}
E_{\rm {DC}}(\textbf R)=\langle \Psi_{\rm DC}|H|\Psi_{\rm DC}\rangle
\end{equation}
is the density-constrained energy.
Since this density-constrained potential still contains the binding energies of individual nuclei which should be subtracted out,
the heavy-ion interaction potential is deduced as
\begin{equation}
V(\textbf R)=E_{\rm {DC}}(\textbf R)-E_{\rm {A1}}-E_{\rm {A2}},
\label{VB}
\end{equation}
where $E_{\rm {A1}}$ and $E_{\rm {A2}}$ are the binding energies of the two individual nuclei.
One should note that this procedure does not affect the TDHF time evolution and contains no free parameters or normalization.

\subsection{Bare potential from FHF approach}

In previous subsection, the DC-TDHF technique has been introduced to compute the nucleus-nucleus potential in a dynamical microscopic way.
All of the dynamical effects included in TDHF is then directly incorporated in the potential. Here we look for a different approach
to produce a bare potential which does not include any dynamical contribution, since we aim to disentangle the static and dynamical effects
of the tensor force. The bare nucleus-nucleus potential is defined as the interaction potential between the nuclei in their ground states.
In addition, to preserve the consistency with microscopic calculations, it is necessary to compute the potential from the same EDF used in
HF, TDHF, and DC-TDHF calculations. This is possible using the frozen Hartree-Fock (FHF) technique~\cite{Simenel2013_PRC88-064604},
assuming that the densities of the target and
projectile remain constant and equal to their respective ground state densities.
The potential can then be expressed as
\begin{align}
\label{eq:FD}
V_\mathrm{FD}(\textbf R)=E[\rho_\mathrm{1}+\rho_\mathrm{2}](\textbf R)-E[\rho_\mathrm{1}]-E[\rho_\mathrm{2}],
\end{align}
where $\rho_\mathrm{1}$ and $\rho_\mathrm{2}$ are HF ground state densities of the fragments, and $E[\rho_\mathrm{1}+\rho_\mathrm{2}]$ is the same
Skyrme EDF as defined in Eqs.~(\ref{Energy}) and (\ref{EDFH}). In the FHF approach, the Pauli principle between pairs of nucleons belonging to different
collision partners has been neglected. When the overlap between the density distributions is small, the barrier is almost unaffected by the inclusion of the Pauli principle.
However, at larger overlaps of the densities where the Pauli principle is expected to play a more important role, the FHF approximation may not properly
account for the potential, particularly the inner part~\cite{Simenel2017_PRC95-031601}.

\section{results}
\label{discuss}

The concept of using density as a constraint for calculating collective states from TDHF time evolution was first introduced in the mid 1980s~\cite{Cusson1985_ZPA320-475}, and was used for the microscopic description of nuclear molecular resonances~\cite{Umar1985_PRC32-172}.
In recent years, the DC-TDHF approach has demonstrated its feasibility and success in explaining sub-barrier fusion dynamics for a wide range of reactions.
This is rather remarkable given the fact that the only input in DC-TDHF is the Skyrme effective interaction, and there are no adjustable parameters.
In the present work, we focus on how the tensor force affect the nucleus-nucleus potential which is vital for the theoretical analysis of sub-barrier fusion dynamics.
We have thus chosen ten representative reactions with proton and neutron numbers of reaction partners corresponding to the magic numbers 8, 20, 28, 50, and 82,
in which the spin-saturated shells are 8 and 20.

In the numerical simulation the static HF ground state for the reaction partner has been calculated on the symmetry-unrestricted three-dimensional grid.
The resulting Slater determinants for each nucleus comprise the larger Slater determinant describing the colliding system during the dynamical evolution.
The TDHF time propagation is performed using a Taylor-series expansion up to the sixth order of the unitary boost operator with
a time step of $0.2 ~\mathrm {fm}/c$. For the dynamical evolution, we use a numerical box of 48 fm along the collision axis and 24 fm in the
other two directions and a grid spacing of 1.0~fm. The initial separation between the two nuclei is 20~fm. The choice of these
parameters assures good numerical accuracy in the unrestricted TDHF evolution. We have simultaneously performed the density constraint
calculations utilizing the DC-TDHF method at every 10-20 time steps (corresponding to 2-4~$\mathrm {fm}/c$ interval).
The convergence property in DC-TDHF calculations is as good if
not better than in the traditional constrained HF with a constraint on a single collective degree of freedom.

We employ the Skyrme interaction in the calculations, in which the tensor force has been constructed in two ways.
One is to add the force perturbatively to the existing standard interactions, for instance,
the existing Skyrme parameter SLy5~\cite{Chabanat1998_NPA635-231} plus tensor force, denoted as SLy5t~\cite{Colo2007_PLB646-227}.
The comparison between calculations with SLy5 and SLy5t addresses the question on how much of the changes is caused by tensor force itself.
Another approach is to readjust the full set of Skyrme parameters self-consistently.
This strategy has been adopted in Ref.~\cite{Lesinski2007_PRC76-014312} and led to
the set of T$IJ$ parametrizations with a wide range of isoscalar and isovector tensor couplings.
Due to its fitting strategy, the contributions from the tensor force and the rearrangement of all other terms could be physically entangled.
\begin{figure}
\includegraphics[width=8.6cm]{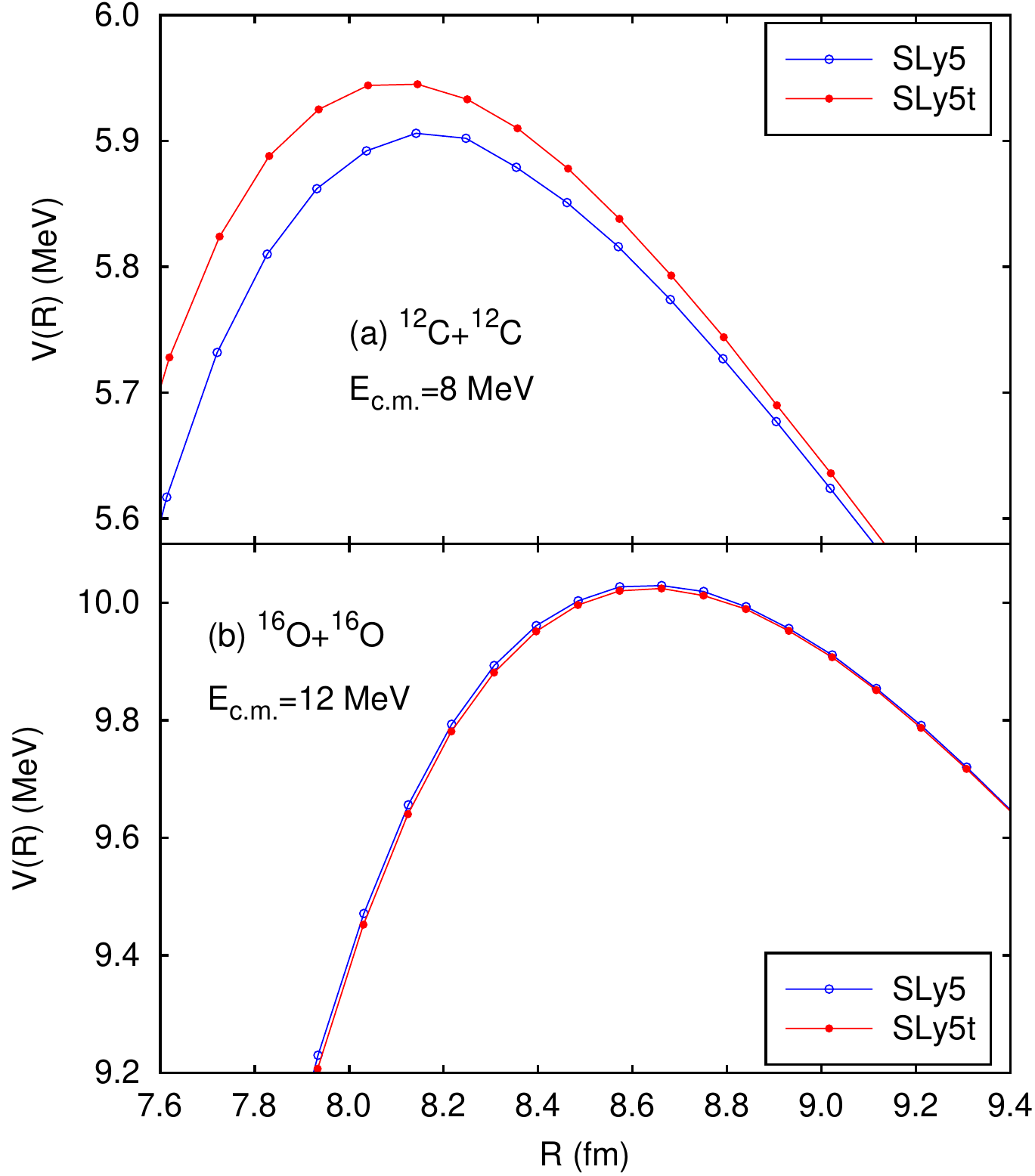}
\caption{(Color online) Internuclear potential obtained from DC-TDHF approach shown for the evolution of the systems (a) $^{12}\mathrm{C}+\mathrm{^{12}C}$
at $E_{\mathrm{c.m.}}=8$ MeV  and (b) $^{16}\mathrm{O}+\mathrm{^{16}O}$ at $E_{\mathrm{c.m.}}=12$ MeV with SLy5 (open circle) and SLy5t (solid circle) forces.
\label{Fig:light}}
\end{figure}

For light systems, we choose the spin-unsaturated $^{12}\mathrm{C}+\mathrm{^{12}C}$ and spin-saturated $^{16}\mathrm{O}+\mathrm{^{16}O}$ reactions
for comparison. As we have reported in Ref.~\cite{Umar2014_PRC89-034611}, the potential barriers are sensitive to the
colliding energy. Hence, the same initial energy, close to the Coulomb barrier, is used for the reaction with and without tensor forces.
In Fig.~\ref{Fig:light}, we plot the ion-ion potentials obtained from Eq.~(\ref{VB}) using the DC-TDHF approach for (a) $^{12}\mathrm{C}+\mathrm{^{12}C}$
at $E_{\mathrm{c.m.}}=8$ MeV  and (b) $^{16}\mathrm{O}+\mathrm{^{16}O}$ at $E_{\mathrm{c.m.}}=12$ MeV with SLy5 (open circle) and SLy5t (solid circle) forces.
Both nuclei, $^{12}\mathrm{C}$ and $\mathrm{^{16}O}$, show spherical ground states with tensor (SLy5t) and without tensor (SLy5) forces, which
are in agreement with experimental data and other calculations. The correct description of the initial shape of
target and projectile nucleus is important for the dynamical evolution of heavy-ion collisions.
We see that for the spin-unsaturated system $^{12}\mathrm{C}+\mathrm{^{12}C}$, the potential with the tensor force included has an overall higher
interaction barrier than without the tensor force, although the difference of the potential barrier peak is small at roughly 0.07~MeV.
For the spin-saturated system $^{16}\mathrm{O}+\mathrm{^{16}O}$, the internuclear potential is close with and without tensor force,
having a barrier height of 10.02~MeV and a peak location of 8.66~fm. This indicates that the tensor force has negligible effect on the near-barrier fusion
for the spin-saturated system $^{16}\mathrm{O}+\mathrm{^{16}O}$, which is consistent with the findings in Ref.~\cite{Stevenson2016_PRC93-054617}.
For these light systems the tensor force shows a small effect on the interaction potential.
\begin{figure*}
\includegraphics[width=15.7cm]{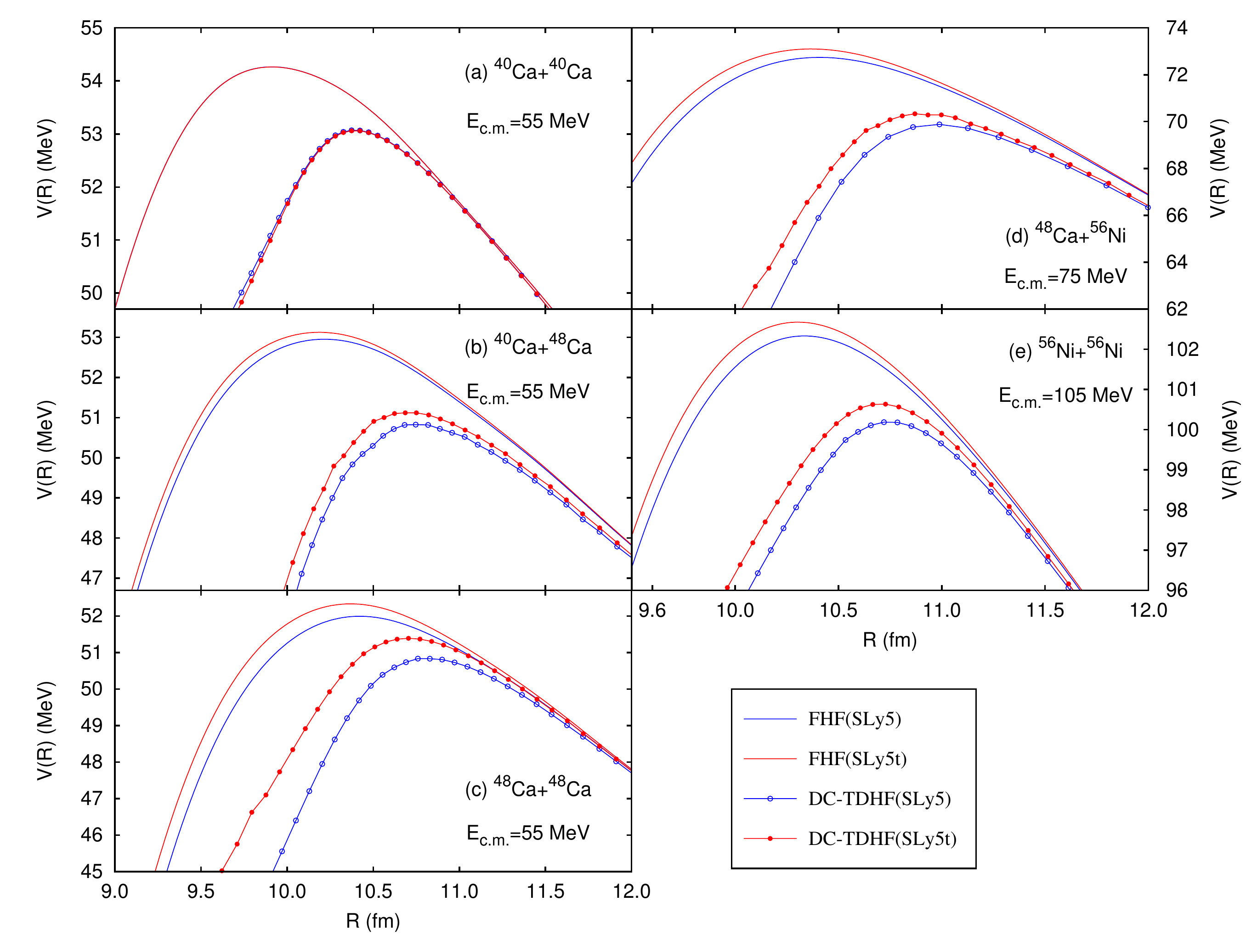}
\caption{(Color online) Internuclear potential obtained from FHF and DC-TDHF approaches for the Ca+Ca, Ca+Ni, and Ni+Ni reactions
with tensor (SLy5t) and without tensor (SLy5) forces.
\label{Fig:medium}}
\end{figure*}

For reactions involving two medium mass nuclei, we have chosen five representative reactions $^{40}\mathrm{Ca}+\mathrm{^{40}Ca}$,
$^{40}\mathrm{Ca}+\mathrm{^{48}Ca}$, $^{48}\mathrm{Ca}+\mathrm{^{48}Ca}$, $^{48}\mathrm{Ca}+\mathrm{^{56}Ni}$, and $^{56}\mathrm{Ni}+\mathrm{^{56}Ni}$,
which vary by the total number of spin-unsaturated magic numbers in target and projectile by 0, 1, 2, 3, and 4.
In these collisions, the reaction partners are closed-shell corresponding to 20 (spin-saturated) and 28 (spin-unsaturated) neutron or proton magic numbers.
To disentangle the static (e.g. modification of ground-state density) and dynamical (e.g. modification of couplings, dissipation, and transfer) origins of the tensor
force, the nucleus-nucleus potentials obtained both from FHF and DC-TDHF calculations are shown in Fig.~\ref{Fig:medium} for all Ca and Ni reactions.
In the initial state of the collision dynamics, the deviation of the static FHF potential from the dynamical DC-TDHF result is the order of smaller than 10 keV.
For all the Ca and Ni reactions, we observe that the nucleus-nucleus potentials are considerably different for the static FHF and dynamical DC-TDHF results.
The static potentials are systematically higher than the dynamical results, and the barrier peaks are located at smaller relative distance with FHF. In particular,
the inner part of the potential, having strong effect on the sub-barrier fusion, presents more significant difference for FHF and DC-TDHF results.
This behavior is well understood and is a consequence of the absence of Pauli principle and excitations for the frozen density overlaps in FHF
potentials~\cite{Simenel2013_PRC88-064604,Guo2018_PLB782-401,Simenel2017_PRC95-031601}, thus the difference between FHF and DC-TDHF is due to dynamical effects.
Another interesting observation is that the variation of dynamical barriers due to tensor force is systematically greater than the
ones for the static barriers.
This indicates that the tensor force influences not only
the ground-state single-particle levels, but also the dynamical effects including nucleon transfer, the couplings to low-lying states, and intrinsic
excitations.
In Ref.~\cite{Guo2018_PLB782-401}, how these dynamical effects affect the fusion barriers heights, computed directly from TDHF, have been investigated to study the
role of tensor force on above-barrier fusion dynamics.
We note that in Ref.~\cite{Guo2018_PLB782-401}, for the $^{48}\mathrm{Ca}+\mathrm{^{56}Ni}$ system, the
tensor force was observed to decrease the barrier height in direct TDHF calculations, which is the opposite of
the trend observed here.
This difference might arise from the dynamical energy-dependent effects introduced by the tensor force that
are not captured by the DC-TDHF potential.
\begin{figure}[t]
\includegraphics[width=8.6cm]{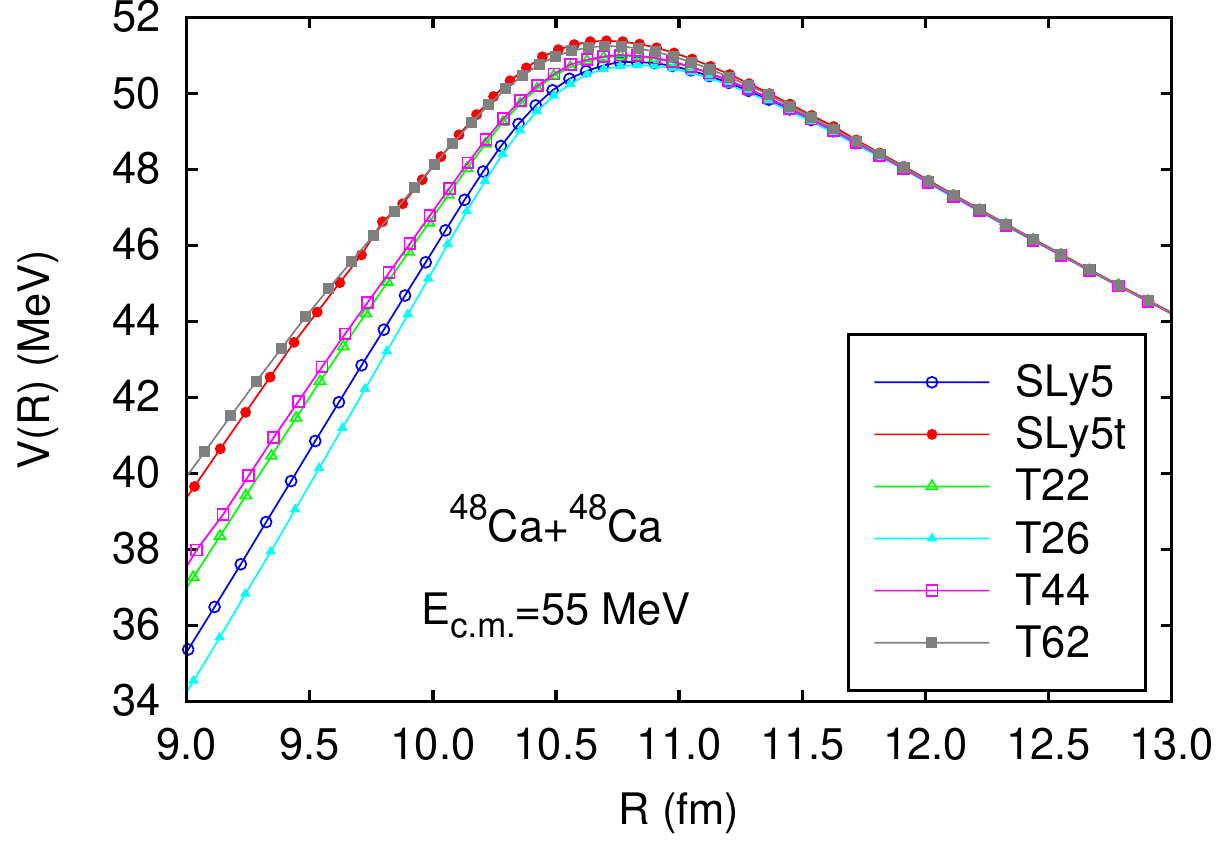}
\caption{(Color online) Internuclear potential obtained from DC-TDHF approach for the reaction $^{48}\mathrm{Ca}+\mathrm{^{48}Ca}$ with SLy5, SLy5t,
T22, T26, T44, and T62 forces.
\label{Fig:TIJ}}
\end{figure}

For the spin-saturated reaction $^{40}\mathrm{Ca}+\mathrm{^{40}Ca}$, the interaction potential remains nearly unchanged by the inclusion of tensor force
for both static and dynamical cases, indicating that the tensor force has almost no impact on the dynamical evolution for spin-saturated systems,
since the contribution of tensor force is expected to be nearly zero for the ground state of spin-saturated nuclei.
For the spin-unsaturated reactions, the barriers with tensor force SLy5t are systematically higher than those without the tensor force SLy5.
This indicates a fusion hindrance effect due to the tensor force in this mass region. Empirically, 1 MeV larger in the inner part of the potential barrier
can cause one order lower in the fusion cross section at sub-barrier energies. From the comparison of dynamical potentials for SLy5 and SLy5t,
the potential barrier increases from a fraction up to a few MeV due to tensor force, which may results in changes of the sub-barrier fusion cross sections by a few orders of magnitude. For the medium mass systems
with proton or neutron magic shells 20 and 28, the tensor force has a significant effect on the nucleus-nucleus potential, particularly in the inner region.

Until now, the studies have utilized the tensor force SLy5t.
To obtain a comprehensive and rigorous understanding of the effects of the tensor force in heavy-ion collisions,
we now proceed to a comparison among the results of various forces, for which the coupling constants
are listed in Tab.~\ref{tab:CC}. Taking the reaction $^{48}\mathrm{Ca}+\mathrm{^{48}Ca}$ as an example,
\begin{table}[!hbt]
    \caption{Isoscalar and isovector spin-current coupling constants in units of MeV~fm$^5$.}
    \label{tab:CC}
    \begin{ruledtabular}
        %\begin{tabular*}{0.4\textwidth}{@{\extracolsep{\fill}}rrr}
        \begin{tabular}{c c c}
            Force & $\mathrm{C}^\mathrm{J}_0$ & $\mathrm{C}^\mathrm{J}_1$  \\
            \hline
            T22   &     0   &      0   \\
            T26   &   120   &    120   \\
            T44   &   120   &      0   \\
            T62   &   120   &   -120   \\
            SLy5  &  15.65  &    64.55  \\
            SLy5t & -19.35  &   -70.45
        \end{tabular}
    \end{ruledtabular}
\end{table}
we show the nucleus-nucleus potential with the six forces SLy5, SLy5t T22, T26, T44, and T62 in Fig.~\ref{Fig:TIJ}.
For T$22$ and T$44$ the potentials are close to each other, indicating the isoscalar tensor coupling has
negligible effect in this reaction. By comparing the results with T$26$, T$44$, and T$62$, the potential increases as the isovector tensor
coupling decreases. This clear dependence of isoscalar and isovector tensor coupling may be due to the interplay between tensor terms
and rearrangement of mean-field. The effect of the isoscalar tensor with the proton and neutron single particle spectrum moving
in the same way seems to be canceled by the refitting of the parameters.
However, the refitting does not incorporate the effect of isovector tensor in the same way.
Detailed discussions on this can be found in Ref.~\cite{Guo2018_PLB782-401}.
The T$62$ (T$26$) interaction also leads to similar potentials as SLy5t (SLy5), even though they have quite different tensor coupling constants,
because the rearrangement of the mean-field for T$62$ (T$26$)
produce additional effects which cancel part of the tensor force in SLy5t (SLy5).
\begin{figure}[t]
    \includegraphics[width=8.6cm]{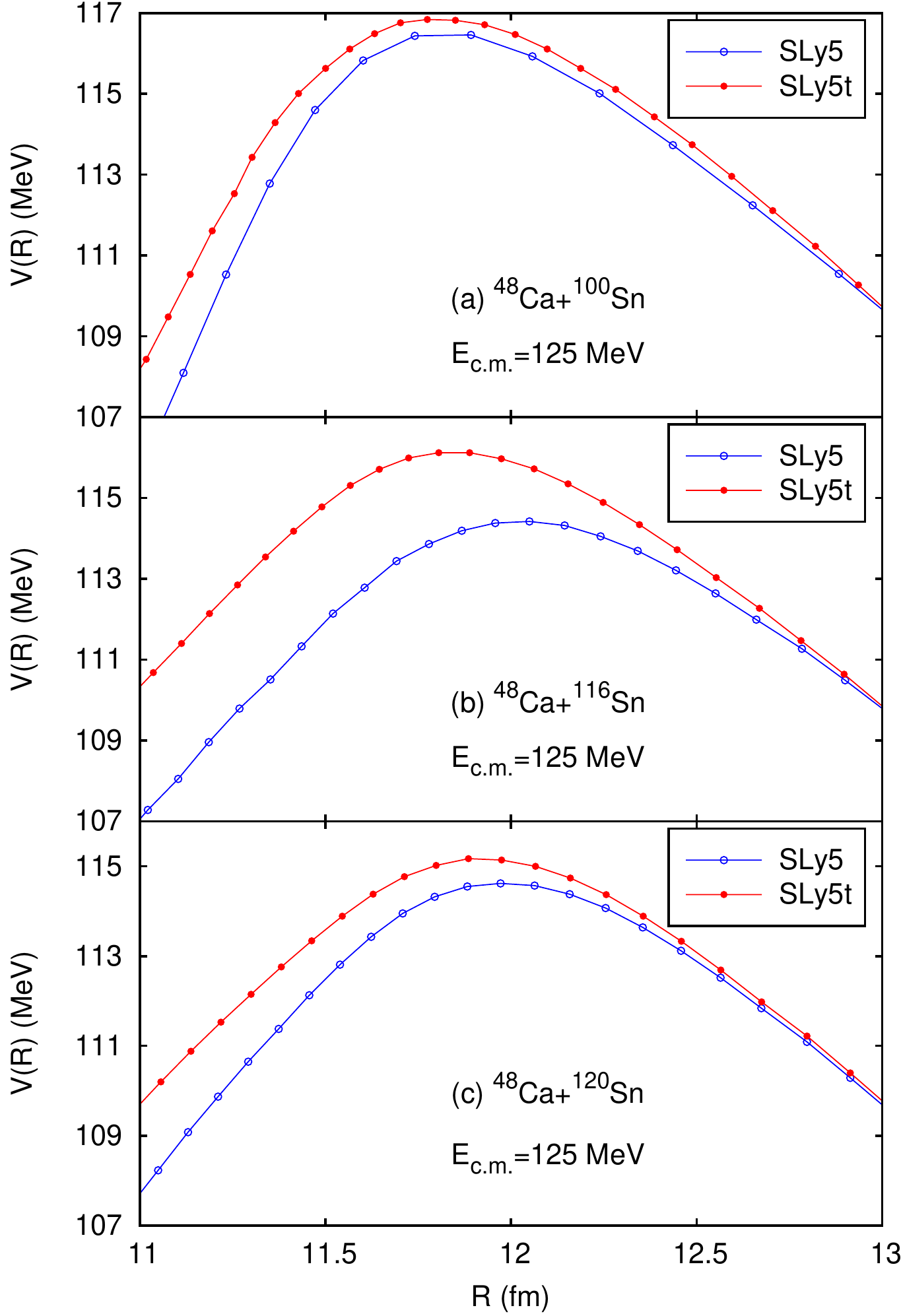}
    \caption{(Color online) Internuclear potential obtained from DC-TDHF approach for the Ca+Sn systems with SLy5 (open circle) and SLy5t (solid circle) forces.
        \label{Fig:heavy}}
\end{figure}

To gain a better insight into the tensor force, the dynamical potential is shown in Fig.~\ref{Fig:heavy} for various Ca+Sn systems
which involve one medium and
one heavy reaction partner. For $^{48}\mathrm{Ca}$, $^{100}\mathrm{Sn}$, and $^{120}\mathrm{Sn}$, the ground states are found to be spherical for
both SLy5 and SLy5t. However, the $^{116}\mathrm{Sn}$ nucleus exhibits small quadrupole deformation $\beta_2$ of 0.077 and 0.026 for SLy5 and SLy5t, respectively,
for which the deformation difference arises from the tensor force. Since the outcome of collision dynamics strongly depends on the deformation orientation
of colliding partners, the deformed nucleus $^{116}$Sn is initially set as the tip orientation in both SLy5 and SLy5t with the symmetry axis of $^{116}$Sn parallel to the internuclear axis. We find that, for the Ca+Sn systems,
the effects of the tensor force show similar trends as in the spin-unsaturated Ca+Ca, Ca+Ni, and Ni+Ni systems presented in Fig.~\ref{Fig:medium}.
The tensor force has the largest effect on the reaction $^{48}\mathrm{Ca}$+$^{116}\mathrm{Sn}$ as compared to the reactions with $^{48}\mathrm{Ca}$ colliding $^{100}\mathrm{Sn}$ and $^{120}\mathrm{Sn}$
isotopes, which may be due to the strong effect of the tensor force on the energy difference of single-proton states  $1\mathrm{h}_{11/2}$ and $1\mathrm{g}_{9/2}$ along the Z=50 isotopes for $^{116}\mathrm{Sn}$, as shown in Ref.~\cite{Colo2007_PLB646-227}.
Another suspected cause for this large effect arising from the tensor force in the $^{48}\mathrm{Ca}$+$^{116}\mathrm{Sn}$ reaction is the static deformation effects leading to a vastly different dynamical path for the system.

\section{Summary}
We incorporate the full tensor force into the FHF and DC-TDHF approaches to investigate the impact of the tensor force on heavy-ion internuclear potentials for ten representative systems in different mass regimes.
As expected we find that static potentials are systematically higher than the dynamical results, however, the variation of dynamical potential barriers induced by tensor force
is larger than those of the static case, which are attributed to the microscopic dynamical effects included in TDHF.
For light systems, the tensor force is found to have small effects on the nucleus-nucleus potential, with the barrier height and inner part of the barrier changing by a fraction of an MeV. Even this small change may lead to large effects in cross sections when considering deep sub-barrier collisions at energy scales common in astrophysical systems.
For medium and heavy spin-unsaturated reactions the effect is much more pronounced, with changes from a fraction of an MeV to almost 2~MeV for the barrier height. These differences indicate an important impact on sub-barrier fusion dynamics and a substantial fusion hindrance effect arising from the tensor force.

The fully microscopic TDHF theory has shown itself to be rich in
nuclear phenomena and continues to stimulate our understanding of nuclear dynamics.
The time-dependent mean-field studies seem to show that the dynamic evolution
builds up correlations that are not present in the static theory.
While modern Skyrme forces provide a much better description of static nuclear properties
in comparison to the earlier parametrizations, there is a need to obtain even better
parametrizations that incorporate deformation and reaction data into the fit process.
The tensor force should be a part of these investigations.
\label{summary}

\section{Acknowledgments}
%todo add acknowledgments
This work is partly supported by NSF of China (Grants No. 11175252 and 11575189),
NSFC-JSPS International Cooperation Program (Grant No. 11711540016), and Presidential Fund of UCAS,
and by the U.S. Department of Energy under grant No. DE-SC0013847.
The computations in present work have been performed on the High-performance Computing Clusters of SKLTP/ITP-CAS and
Tianhe-1A supercomputer located in the Chinese National Supercomputer Center in Tianjin.

\bibliography{ref}
\end{document}